\documentclass{svproc}
\usepackage{url}

\usepackage{graphicx}
\usepackage{listings}
\usepackage{color}
\definecolor{javared}{rgb}{0.6,0,0} 
\definecolor{javagreen}{rgb}{0.25,0.5,0.35} 
\definecolor{javapurple}{rgb}{0.5,0,0.35} 
\definecolor{javadocblue}{rgb}{0.25,0.35,0.75} 
\usepackage{soul}
\begin{document}
\mainmatter              
\title{Open-Ended Automatic Programming Through Combinatorial Evolution}
\titlerunning{Open-Ended Automatic Programming Through Combinatorial Evolution}  
%
\author{Sebastian Fix \and
Thomas Probst \and
Oliver Ruggli \and \\
Thomas Hanne \and
Patrik Christen}
\authorrunning{S. Fix et al.} 
%
%
\institute{FHNW University of Applied Sciences and Arts Northwestern Switzerland,\\ 4600 Olten, Switzerland\\
\email{patrik.christen@fhnw.ch}
}

\maketitle              

\begin{abstract}
Combinatorial evolution -- the creation of new things through the combination of existing things -- can be a powerful way to evolve rather than design technical objects such as electronic circuits. Intriguingly, this seems to be an ongoing and thus open-ended process creating novelty with increasing complexity. Here, we employ combinatorial evolution in software development. While current approaches such as genetic programming are efficient in solving particular problems, they all converge towards a solution and do not create anything new anymore afterwards. Combinatorial evolution of complex systems such as languages and technology are considered open-ended. Therefore, open-ended automatic programming might be possible through combinatorial evolution. We implemented a computer program simulating combinatorial evolution of code blocks stored in a database to make them available for combining. Automatic programming in the sense of algorithm-based code generation is achieved by evaluating regular expressions. We found that reserved keywords of a programming language are suitable for defining the basic code blocks at the beginning of the simulation. We also found that placeholders can be used to combine code blocks and that code complexity can be described in terms of the importance to the programming language. As in a previous combinatorial evolution simulation of electronic circuits, complexity increased from simple keywords and special characters to more complex variable declarations, class definitions, methods, and classes containing methods and variable declarations. Combinatorial evolution, therefore, seems to be a promising approach for open-ended automatic programming.
\end{abstract}
\section{Introduction}
Genetic algorithms and evolutionary computation in general are widely used for solving optimisation problems \cite{Baeck.1996}. Such algorithms follow the paradigm of biological evolution. They consist of a collection of virtual organisms, where every organism represents a possible solution to a given problem. Some fitness measure is then calculated for each organism in an iterative process and it tries to find improved solutions by forming random mutations and crossovers on them.\\
In contrast to such evolutionary computation, \emph{combinatorial evolution} as proposed by W. Brian Arthur \cite{Arthur.2009,ArthurSydney.2018}, makes no modifications to the organisms themselves. New solutions are formed through the combination of existing components which then form new solutions in later iterations with the goal of satisfying certain needs. The more useful a combination is, the higher is its need rating. Combining existing components to construct new components can be observed in the evolution of technology \cite{Arthur.2009,ArthurSydney.2018}. For instance, the invention of radar was only possible through combining simpler electronic parts fulfilling functions like amplification and wave generation \cite{Arthur.2006}. In order to investigate combinatorial evolution, Arthur and Polak \cite{Arthur.2006} created a simple computer simulation, where electronic circuits were evolved in a combinatorial manner. Their simulation started by randomly combining primitive elementary logic gates and then used these simpler combinations for more complicated combinations in later iterations. Over time, a small number of simple building blocks was transformed into many complicated ones, where some of them might be useful for future applications. It was concluded that combinatorial evolution allows building some kind of library of building blocks for the creation of future and more complicated building blocks.\\
Intriguingly, combinatorial evolution is a key ingredient to achieve open-ended evolution \cite{Thurner.2011,Thurner.2018}, that is the ongoing creation of novelty \cite{Banzhaf.2016,Taylor.2019}. This contrasts classical computational approaches where the aim is to converge towards a solution as fast as possible. Computational approaches according to open-ended evolution are therefore not more efficient but they are more creative since they generate ongoing novelty.\\
Here we want to explore whether combinatorial evolution could be also applied to software development, more specifically to automatic programming to eventually make it open-ended \cite{Christen.2021}. An early idea of automatic programming was to implement high-level programming languages that are more human readable resulting in compilers, which produce low-level programs -- down to machine code -- from human readable syntax \cite{Chun.2005}. However, human input in some form was still needed and the programming task was simply transferred to a higher level. Furthermore, the software solution is limited by the programmer's capabilities and creativity. Language therefore remains a barrier between programmers and computers. A way around this barrier would be to let the computer do the programming (also occasionally denoted as metaprogramming \cite{Czarnecki.2000}), which might even lead to better programs. Koza \cite{Koza.1994} addressed this issue through genetic programming, where populations of computer programs are generated by a computer using genetic algorithms. The problem space consists of programs that try to solve (or approximately solve) problems. It has been demonstrated that random mutations and crossovers in source code can effectively contribute in creating new sophisticated programs \cite{poli2007genetic}.\\
Therefore, it seems possible to define a programming task and let a computer do the programming. However, looking at the process of software development, programming seems more comparable to technological rather than biological evolution. Existing libraries or algorithms are often integrated into new software without the necessity of modifying them. Therefore, an automatic programming approach that creates new computer programs by means of combinatorial evolution might be an interesting alternative to genetic programming. Also, due to open-endedness, combinatorial evolution holds the promise to be more creative generating ongoing novelty. In the present study we investigate ways to define a programming task for automatic programming through combinatorial evolution including the evaluation of the generated code with a need rating. Our research question is whether it is possible to generate computer programs of increasing complexity using automatic programming through combinatorial evolution. Specifically, we ask what kind of basic code blocks are needed at the beginning? How are these code blocks implemented to allow them to combine? How can code complexity be measured?
\section{Automatic Programming}
Since the development of computers, it has been a challenge to optimise and adapt program code to access the potential performance of a computer. While the computational power of computers has been steadily increasing in recent years, program code is still limited by the ability of programmers to create efficient and functioning code. Programming languages have also evolved over the past decades. The development of programming languages has sought to provide programmers with abstractions at higher levels. However, this also led to limitations, especially regarding performance and creativity. It is thus intriguing to shift the programming to the computer itself. Most of the programming is currently done by human programmers, which often leads to a time-intensive and error-prone process of software development. The idea that computers automatically create software programs has been a long-standing goal \cite{Becker.2017} with the potential to streamline and improve software development.\\
Automatic programming was first considered in the 1940s describing the automation of a manual process in general and with the goal to maximise efficiency \cite{Parnas.1985}. Later, automatic programming was considered a type of computer programming in which code is generated using tools that allow developers to write code at a higher level of abstraction \cite{Parnas.1985}. There are two main types of automatic programming: \textit{application generators} and \textit{generative programming}. Cleaveland \cite{Cleaveland.1988} describes the development of application generators as the use of high-level programming models or templates to translate certain components into low-level source code. Generative programming, on the other hand, assists developers in writing programs. This can be achieved, e.g. by providing standard libraries as a form of reusable code \cite{Czarnecki.2000}. In generative programming it is crucial to have a domain model, which consists of three main parts: a problem space, a solution space, and a configuration knowledge mapping that connects them \cite{RN111}. The problem space includes the features and concepts used by application engineers to express their needs. These can be textual or graphical programming languages, interactive wizards, or graphical user interfaces. The solution space consists of elementary components with a maximum of combinability and a minimum of redundancy. The configuration knowledge mapping presents a form of generator that translates the objects from the problem space to build components in the solution space \cite{Czarnecki.2000}. Most recently, automatic programming shifted towards higher level programming languages and incorporating even more abstraction \cite{RN24}.\\
While these kinds of automatic programming heavily depend on human interaction and thus the capabilities and creativity of programmers, genetic programming can be regarded an attempts to reduce this dependency and shift the focus to automation done by the computer itself. Koza \cite{Koza.1994} describes genetic programming as a type of programming in which programs are regarded as genes that can be evolved using genetic algorithms \cite{Holland.1992,Holland.2012}. It aims to improve the performance of a program to perform a predefined task. According to Becker et al. \cite{Becker.2017}, a genetic algorithm takes, as an input, a set of instructions or actions that are regarded as genes. A random set of these instructions is then selected to form an initial sequence of DNA. The whole genome is then executed as a program and the results are scored in terms of how well the program solves a predefined task. Afterwards, the top scorers are used to create offspring, which are rated again until the desired program is produced. To find new solutions, evolutionary techniques such as crossover, mutation, and replication are used \cite{Pillay.2007}. Crossover children are created by picking two parents and switching certain components. Another technique is mutation, which uses only one individual parent and randomly modifies its parts to create a new child. Sometimes parents with great fitness will be transferred to the next iteration without any mutation or crossover because they might do well in later steps as well.
\section{Combinatorial Evolution}
With combinatorial evolution, new solutions build on combinations of previously discovered solutions. Every evolution starts with some primitive, existing building blocks and uses them to build combinations. Those combinations are then stored in an active repertoire. If the output satisfies a need better than an earlier solution, it replaces the old one and will be used as the building block in later iterations. Building blocks are thus not modified, they are combined together creating new building blocks. The result is a library of functionalities that may be useful for a solution in the future \cite{Arthur.2009,ArthurSydney.2018}.\\
As Ogburn \cite{Ogburn.1922} suggested, the more equipment there is within a material culture, the greater the number of inventions are. This is known as the Ogburn’s Claim. It can therefore be inferred that the number and diversity of developed components as well as their technological developments matters because next generation components build upon the technological level of the previous, existing components. To investigate this, Arthur and Polak \cite{Arthur.2006} created a simple computer simulation to ‘discover’ new electronic circuits. In their simulation, they used a predefined list of truth tables of basic logic functions such as full adders or n-bit adders. Every randomly created combination represented a potential satisfaction of a need, which was then tested against this list. If the truth table of a newly created circuit matched one from the predefined list, it is added to the active repertoire as it fulfils the pre-specified functionality. Sometimes, it also replaced one that was found earlier, if it used fewer parts and therefore would cost less. New technologies in the real world are not usually found by randomly combining existing ones nor do they exist in a pre-specified list to be compared against. Nevertheless, their needs are generally clearly visible in economics and current technologies \cite{Arthur.2006}.\\
Combinatorial evolution is in general an important element of evolutionary systems. Stefan Thurner and his colleagues developed a general model of evolutionary dynamics in which the combination of existing entities to create new entities plays a central role \cite{Thurner.2011,ThurnerSydney.2018,Thurner.2018}. They were able to validate this model using world trade data \cite{Klimek.2012}, therefore underlining the importance of evolutionary dynamics in economic modelling in general and combinatorial interactions in particular. The model shows punctuated equilibria that are typical for open-ended evolutionary systems \cite{Thurner.2011,ThurnerSydney.2018,Thurner.2018}.
\section{Code Complexity}
Genetic algorithms have been used for automatic programming already, however, a large number of iterations are required to significantly increase code complexity in order to solve more complex problems \cite{harter2019advanced}. It therefore seems beneficial to use combinatorial evolution in which complexity seems to increase in fewer steps and thus less time.\\
Code complexity has been measured in this context with different approaches. The cyclomatic complexity of a code is the number of linearly independent paths within it \cite{ebert2016cyclomatic}.
For instance, if the code contains no control flow elements (conditionals), the complexity would be 1, since there would be only a single path through the code \cite{RN112}. If the code has one single-condition IF statement, the complexity would be 2 because there would be two paths through the code -- one where the IF statement evaluates to TRUE and another one where it evaluates to FALSE \cite{RN112}. Two nested single-condition IFs (or one IF with two conditions) would produce a complexity of 3 \cite{McCabe.1976,RN112}. According to Garg \cite{Garg.2014}, cyclomatic complexity is one of the most used and renowned software metrics together with other proposed and researched metrics, such as the number of lines of code and the Halstead measure. Although cyclomatic complexity is very popular, it is difficult to calculate for object-oriented code \cite{sarwar2013cyclomatic}.
\section{Methods}
\subsection{Development Setup and Environment}
We used the programming language Java though other programming languages would have been feasible as well. The development environment was installed on VirtualBox -- an open source virtualisation environment from Oracle. Oracle Java SE Development Kit 11 was used with Apache Maven as build automation tool. To map the existing code with a database, Hibernate ORM was used. It allows mapping object-oriented Java code to a relational database. Furthermore, code versioning with GitHub was used.
\subsection{Simulation}
Simulations are initialised by adding some basic code building blocks into a repository. The first simulation iteration then starts by randomly selecting code blocks from this repository. Selected blocks are then combined into a new code block, which subsequently gets analysed for its usefulness and complexity. Based on this analysis, the code block is assigned a value. Nonsense code, which is the most common result when randomly combining keywords of a programming language, are assigned a value of 0 and not used any further. Only code blocks with a value greater than 0 are added to the repository and consequently have a chance of being selected in a later iteration.
\subsection{Code Building Blocks}
Preliminary experiments in which code snippets with placeholders were predefined showed that this approach would limit the creativity and complexity of the automatic programming solution by the predefined snippets. The simulation would only create program logic that is already given by the basic set of code blocks.\\
To overcome this limitation, we defined basic code building blocks according to keywords and special characters of the Java programming language, e.g. the keywords \lstinline{int}, \lstinline{for}, \lstinline{class}, and \lstinline{String} as well as the special characters \lstinline{&}, \lstinline{=}, \lstinline{;}, and \lstinline{{}. Additionally, we defined three more extra code blocks: First, \lstinline{PLACEHOLDER} to define where blocks allow other code blocks to be combined and integrated. This is particularly important for nesting certain code elements, such as methods that must be nested into a class construct to be valid Java code. Second, \lstinline{NAME} to name something, e.g. classes, methods, and variables. And third, the special keyword \lstinline{main} in the main method definition.
\subsection{Selecting and Combining Code Blocks}
During the selection process, new source code is generated based on combinations of existing code blocks from the repository. The chance that a particular code block is selected depends on its classification value (see next section). In a first step, a helper function defines a random value of how many code blocks are taken into consideration in the current iteration. There is a minimum of two code blocks required to generate a new code block. The maximum number can be predefined in the program. Arthur and Polak \cite{Arthur.2006} combined up to 12 building blocks. To reduce the number of iterations needed for receiving valid Java code, a maximum of eight blocks turned out to be a good limit. After randomly defining the number of code blocks to be combined, the weighted random selection of code blocks based on their classification value follows. Instead of simply chaining all selected code blocks together, there is also the possibility to nest them into a placeholder if available. A random function decides whether a code block is nested into the placeholder, or simply added to the whole code block. This procedure is important because program code usually exhibits such nested structures.
\subsection{Code Analysis and Building Block Classification}
After the selection and combination process, the newly generated source code is passed into the classification function where it gets analysed.
The classification process is required to weight the different code blocks according to their relevance in the Java programming language and to see whether the code evolved with respect to complexity. This is achieved with regular expression patterns, which allow identifying relevant Java code structures such as classes and methods that can be weighted with predefined classification values for these code structures. Basic structures such as variable declarations are assigned a value of 1. More elaborate structures such as classes have a value of 2 and even more complicated structures such as methods have a value of 3. If a structure contains several of these substructures, their classification values is added. An important structure in many programming languages is the declaration of a variable. With the following regular expression, any declaration of the value types \lstinline{boolean}, \lstinline{byte}, \lstinline{char}, \lstinline{double}, \lstinline{float}, \lstinline{int}, \lstinline{long}, and \lstinline{short} are detected:
\begin{lstlisting}[language=Java]
(PLACEHOLDER(?!PLACEHOLDER))?
(boolean|byte|char|double|float|int|long|short) NAME;
(PLACEHOLDER(?!PLACEHOLDER))?
\end{lstlisting}
Other important elements are brackets. E.g. they are used in methods and classes specifying the body. The syntax is given by the programming language. Placeholders inside brackets are important, they allow new code to be injected into existing code blocks in future combinations. We therefore created the following regular expression:
\begin{lstlisting}[language=Java]
^(\{PLACEHOLDER\}|\(PLACEHOLDER\))$
\end{lstlisting}
As already shown in the simple simulation with electronic circuits \cite{Arthur.2006}, one needs a minimal complexity of the initial building blocks to be able to generate useful and more complex future combinations. Classes and methods are essential to build anything complex in Java. Therefore, regular expressions were implemented to identify valid classes and methods. Valid means, the element is closed and it successfully compiles. Variable declarations and methods are allowed to be nested in the class structure. The following regular expression to detect classes was developed:
\begin{lstlisting}[language=Java]
(protected|private|public) class NAME \{
((boolean|byte|char|double|float|int|long|short) NAME; 
|(protected|private|public) void NAME\(
((boolean|byte|char|double|float|int|long|short) NAME)?\) \{
((boolean|byte|char|double|float|int|long|short) NAME; 
|PLACEHOLDER(?!PLACEHOLDER))*\} 
|PLACEHOLDER(?!PLACEHOLDER))*\}$
\end{lstlisting}
A valid method needs to be correctly closed and can contain either a placeholder or a variable declaration. 
The following regular expression to detect methods was developed:
\begin{lstlisting}[language=Java]
(PLACEHOLDER(?!PLACEHOLDER))?
(protected|private|public) void NAME\(
((boolean|byte|char|double|float|int|long|short) NAME)?\) \{
((boolean|byte|char|double|float|int|long|short) NAME; 
|PLACEHOLDER(?!PLACEHOLDER))*\}
(PLACEHOLDER(?!PLACEHOLDER))?
\end{lstlisting}
\subsection{Regular Expression Validation}
In some preliminary experiments, we automatically compiled source code files of newly combined code blocks to check whether they are valid. However, this process is too time consuming to allow large numbers of iterations. An iteration required one to three seconds compilation time. As combinatorial evolution relies on rather large numbers of iterations, we instead used regular expressions to check whether newly combined code blocks compile and are thus valid. Java allows compiling regular expression into a pattern object, which can then be used to match it with a string containing the code to be tested. It turned out to be a much faster alternative to the actual compilation of source code files.
\section{Results}
Using Java keywords for the initial basic code blocks, we found the first useful combinations of code blocks within 100'000 iterations in a simulation of 1.6 billion iterations, which took approximately 5 hours on a desktop computer. These code blocks mainly consisted of combinations of three basic building blocks classified with a value of 1. Table~1 shows some examples that were found in the simulation. Such combinations are typically assigned a small classification value due to their simplicity, keeping in mind that only code blocks that are assigned values greater than 0 are added to the code block repository for later combinations.
%
\begin{table}
\caption{Examples of newly generated code blocks within the first 100'000 iterations of a combinatorial evolution simulation. Class refers to the classification value representing how useful the code block is in programming.}
\begin{center}
\begin{tabular}{rclc}\hline
Iteration & Block & New Code Block & Class\\
\hline
4'647 & 25 & \lstinline$short NAME ;$ & 1\\
\hline
16’394 & 30 & \lstinline$public void NAME$ & 1 \\
\hline
22’729 & 34 & \lstinline$boolean NAME ;$ & 1 \\
\hline
50’419 & 42 & \lstinline$protected class NAME$ & 1 \\
\hline
58’595 & 44 & \lstinline${ PLACEHOLDER }$ & 1 \\
\hline
93’722 & 55 & \lstinline$public class NAME$ & 1 \\
\hline
\end{tabular}
\end{center}
\end{table}
\begin{table}
\caption{Examples of newly generated code blocks within a wide range of iterations of a combinatorial evolution simulation of 1.6 billion iterations. Class refers to the classification value representing how useful the code block in programming.}
\begin{center}
\begin{tabular}{rclc}
\hline
Iteration & Block & New Code Block & Class\\
\hline
58’903 & 45 & \begin{lstlisting}[language=Java]
protected class NAME { 
  PLACEHOLDER 
}
\end{lstlisting} & 2\\
\hline
112’609 & 61 & \begin{lstlisting}[language=Java]
public class NAME { 
  PLACEHOLDER 
}
\end{lstlisting} & 2 \\
\hline
$>1\cdot10^{9}$ & 168 & \begin{lstlisting}[language=Java]
public void NAME ( ) { 
  PLACEHOLDER 
}
\end{lstlisting} & 3 \\
\hline
$>1\cdot10^{9}$ & 169 & \begin{lstlisting}[language=Java]
public class NAME { 
  public void NAME ( ) { 
    PLACEHOLDER 
  } 
  short NAME ; 
}
\end{lstlisting} & 6 \\
\hline
$>1\cdot10^{9}$ & 170 & \begin{lstlisting}[language=Java]
protected class NAME { 
  boolean NAME ; 
  public void NAME ( ) { 
    PLACEHOLDER 
  } 
}
\end{lstlisting} & 6 \\
\hline
\end{tabular}
\end{center}
\end{table}
It did not take long for the combinatorial evolution simulation to find the first combinations that consisted of previously found code blocks as illustrated in Table~2. E.g. code block 45 – which consists of block 42 and block 44 – was found only 308 iterations later. Though it took some time to find a Java method in code block 168, only a small number of iterations later, many subsequent code blocks followed with higher classification values. Code blocks 169 and 170 characterise Java classes that contain methods and declarations of variables.
%
It took considerably longer to jump to the next higher classification value of 3 as compared to the jump from value 1 to 2. More than $1\cdot10^{9}$ iterations were required to evolve a method with a placeholder in it, classified with a value of 3. From there it only took a few iterations to jump to classification values of 4, 5, and even 6. Combinations of a method with a variable declaration were assigned a classification value of 4, combinations with a class were assigned a classification value of 5, and combining all three resulted in the assignment of a classification value of 6. 
%
%
%
\section{Discussion and Conclusion}
In the present paper, we investigated whether it is possible to generate computer programs of increasing complexity using automatic programming through combinatorial evolution since this would make it an open-ended process. Specifically, we wanted to know what kind of basic code blocks are needed at the beginning of a simulation, how are these code blocks implemented to allow them to combine, and how can code complexity be measured. To start the first iteration of the combinatorial evolution simulation we needed to define code blocks that existed in the programming framework Java. As initial code blocks we defined reserved keywords of the Java programming language that are used to define classes, methods, initialise variables, and so on. This also includes some special characters used in the programming language that we also added. Placeholders within code blocks are used to allow combining code blocks and thus source code. Newly generated code blocks are assigned a classification value according to their structure, which represents code complexity. The combinatorial evolution simulation generated code blocks including classes, methods, variables, and combinations thereof. It therefore generated code of increasing complexity.\\
Regarding measuring complexity, different approaches to do so, e.g. determining the number of lines of code and McCabe's cyclomatic complexity \cite{McCabe.1976}, have been taken into consideration but the code blocks from the outcomes after nearly 2 billion iterations were in our opinion still too short to implement these complexity measures. Two factors were important why we did not use McCabe's cyclomatic complexity \cite{McCabe.1976}. First, it did not generate the required main method within a reasonable number of iterations, so there was no starting point. Second, we decided to not have the decision code block assigned a value greater than 0 in the initial code blocks. Without any of these code blocks, the complexity would always be evaluated as 1.\\
We conclude that the combinatorial evolution simulation clearly shows how Java code can be automatically created using combinatorial evolution. Simple keywords and special characters were successfully combined into more complex and different structures like variable declarations or methods and in later iterations they even got combined into more sophisticated results such as classes consisting of methods and variable declarations. We also conclude that due to combinatorial evolution, open-ended automatic programming could be achieved, indicating an intriguing approach if creativity is important.\\
The reached limitations of  complexity show that further research is required. Similar observations for genetic programming \cite{harter2019advanced} suggest that more advanced evolutionary operators could be useful. However, already when starting with further elaborated code blocks or when reaching them during previous combinatorial evolution, the goal of automatic programming might come much closer. Therefore, forthcoming research may also study the concept with much increased computational power and distributed computing.

%
%
\bibliographystyle{splncs_srt}
\bibliography{References}
%
%
%
%
%
%
%
%
%
\end{document}